# COMPARISON OF ON-ORBIT MANUAL ATTITUDE CONTROL METHODS FOR NON-DOCKING SPACECRAFT THROUGH VIRTUAL REALITY SIMULATION

Ajit Krishnan[1], Himanshu Vishwakarma[2], Maharudra Kharsade[2], Pradipta Biswas[2]

**[1]Indian Space Research Organization**

**[2]Indian Institute of Science**

**Abstract:** On-orbit manual attitude control of manned spacecraft is accomplished using external visual references and some method of three axis attitude control. All past, present, and developmental spacecraft feature the capability to manually control attitude for de-orbit. National Aeronautics and Space Administration (NASA) spacecraft permit an aircraft windshield type "front view", wherein an arc of the Earth's horizon is visible to the crew in de-orbit attitude. Russian and Chinese spacecraft permit the crew a "bottom view" wherein the entire circular Earth horizon disk is visible to the crew in de-orbit attitude. Our study compared these two types of external views for efficiency in achievement of de-orbit attitude. We used an Unity Virtual Reality (VR) spacecraft simulator that we built in house. The task was to accurately achieve de-orbit attitude while in a 400 km circular orbit. Six military test pilots and six civilians with gaming experience flew the task using two methods of visual reference. Comparison was based on time taken, fuel consumed, cognitive workload assessment and user preference. We used ocular parameters, EEG, NASA TLX and IBM SUS to quantify our results. Our study found that the bottom view was easier to operate for manual de-orbit task. Additionally, we realized that a VR based system can work as a training simulator for manual on-orbit flight path control tasks by astronauts (both pilots and non-pilots). Results from our study can be used for design of manual on-orbit attitude control of present and future spacecrafts.

## 1. Introduction

Outer space provides the advantage of "high ground". Spacecraft operating in this high ground give us obvious benefits in Earth observation, environmental monitoring and global communication. Unobscured by the Earth's atmosphere, spacecraft have deepened our understanding of fundamental scientific principles through the study of planets, stars, galaxies and cosmic phenomena. Our foray into space has led to innovations with varied practical applications on Earth such as satellite navigation, advancement in material sciences, robotics and medical research to name a few. National security has become increasingly reliant on satellite-based surveillance, communication and navigation. The space industry creates jobs, stimulates innovation, entrepreneurship, international collaboration, trade opportunities and fosters the next generation of scientists, engineers and explorers. Space technology stokes our curiosity to explore. The first human crewed spacecraft orbited Earth in 1961 (within 4 years of launching of the first artificial satellite) and man first stepped on the moon in 1969. Since then, space stations such as Salyut, Mir, Skylab, ISS (International Space Station) and Tiangong have built our understanding of sustaining long duration stay in space. For the purpose of this paper, we define manned transport spacecraft as those vehicles that ferry humans from Earth to space and back and are capable of autonomous orbital flight. A complete list of manned transport spacecraft is placed in Table 1.

Table 1 List of Human Spaceflight Programs

| S No | Name | Country | Operational status |
|---|---|---|---|
| 1. | Vostok | USSR | De-commissioned |
| 2. | Mercury | USA | De-commissioned |
| 3. | Voskhod | USSR | De-commissioned |
| 4. | Gemini | USA | De-commissioned |
| 5. | Soyuz | USSR/ Russia | Operational |
| 6. | Apollo | USA | De-commissioned |
| 7. | Space Shuttle | USA | De-commissioned |
| 8. | Crew Dragon | USA | Operational |
| 9. | Shenzhou | China | Operational |
| 10. | CST-100 | USA | Developmental |
| 11. | Orion | USA | Developmental |
| 12. | Dream Chaser | USA | Developmental |

All practically useful spacecraft require the ability to control their flight path. Flight path control implies the ability to determine and change, position and orientation (attitude). Typical flight path control tasks of manned transport spacecraft are as follows:-

- Achievement and maintenance of solar power generating attitude – pointing the electrical power generating side of the spacecraft's solar array towards the Sun.

- Orbit raising/ circularization – Orienting the spacecraft to point the main engine thrust vector along the velocity vector and fire the engines at a predetermined time/ orbital position to increase orbital velocity by a specific amount (delta-V).

- Rendezvous and docking.

- De-orbit – Orienting the spacecraft to point the main engines thrust vector opposite to the velocity vector and fire the engines at a predetermined time/orbital position to decrease orbital velocity by a specific amount (delta-V). This puts the spacecraft into a trajectory that re-enters the Earth's atmosphere.

- Atmospheric flight path control (distinct from the above on-orbit tasks).

Unmanned spacecraft necessarily feature automatic and semi-automatic (inputs from ground) means of flight path control. National Aeronautics and Space Administration (NASA) Human Rating standard for manned spaceflight NPR 8705.2C Para 3.4 (NASA, 2017) mandates manual flight path control by crew, asserting that it is a fundamental element of crew survival. Arguments justifying manual flight control include dealing with unanticipated/ multiple failures, reduced effort required to specify, write and test automation, provide a convenient method of handling common cause software failure and enhanced mission flexibility (Osborn-Hoff, 2022). A number of manned spaceflight missions have been salvaged and crew safely returned to Earth using manual controls, notable examples are Mercury 9, Gemini 4, Gemini 8 and Apollo 13. Manual flight control enables the crew to override faulty automatics, take over the task of failed sensors and/or computers to either accomplish a mission or get back safely. Significant weight reduction is also envisaged through manual flight control by reducing redundant automatic components. For example, while the un-crewed Progress spacecraft feature Sun sensors, the manned Soyuz relies on the crew for the same task and does away with the Sun sensors. All the spacecraft mentioned in Table 1 feature on-orbit manual flight control.

During literature survey presented later in this paper, it was realized that manual flight path control tasks that required using the Earth as visual reference, namely orbit raising/circularization and de-orbit, were done using two distinctly different external visual reference methods. Vostok, Voskhod, Soyuz and Shenzhou permit the crew a "bottom view" wherein the entire circular Earth horizon disk is visible to the crew in de-orbit attitude. A few Mercury missions also provided this bottom view to the crew. However subsequent NASA spacecraft typically provide an aircraft windshield type front view, wherein an arc of the Earth's horizon is visible to the crew in de-orbit attitude. It is evident that front view aids rendezvous and docking due to the typical location of spacecraft docking mechanisms and hence all docking spacecraft provide front view for manual docking. Figure 1 explains these external views diagrammatically.

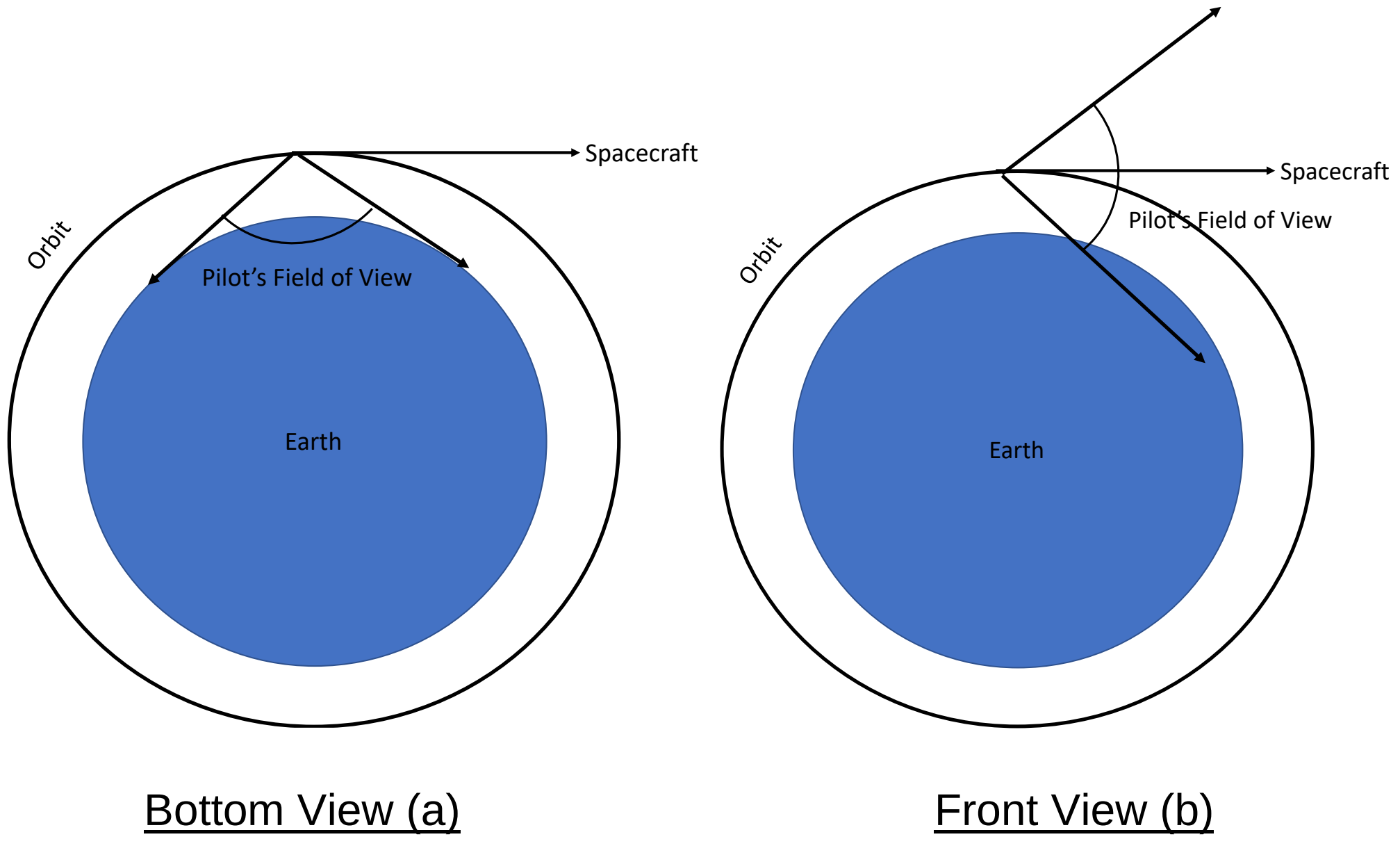

**Figure 1: Types of External Visual References**

This paper aims to present a comparison between the bottom view vs front view for on-orbit attitude control specifically for de-orbit. This study is expected to aid designers of manual control methods for de-orbit maneuver in manned spacecraft for selection of the most suitable method of external visual reference.

## 2. Literature Survey

Salient features of manual flight control implementation in all past, present and developmental manned transport spacecraft have been listed below. Focus is on two aspects – on-orbit tasks and external view. Notable cases of usage which highlight the usefulness of manual flight control have also been mentioned. To appreciate the evolution of current state of the art, spacecrafts have been listed in order of their vintage. While considerable information is available about de-commissioned spacecraft and Soyuz, very limited open-source information is available on operational and developmental spacecraft.

Vostok, the Soviet manned spacecraft, flew the first human being to space. It was a single crew spacecraft. There were 06 crewed Vostok flights from 1961-63. Vostok stayed in low earth orbit after insertion, till it's de-orbit deceleration (retro-fire). Atmospheric descent was uncontrolled (ballistic). Though all flight control tasks were automatic, Vostok crew had the ability to manually control spacecraft orientation in orbit (RSC Energia Publishing House, 1996). This was primarily meant to achieve and maintain de-orbit attitude in case automatic orientation was unsatisfactory (Howard, 1963) and initiate de-orbit (Gerovitch, 2006). A bottom window with an attached periscope type lens system enabled the crew to achieve the desired orientation using the Earth as reference. The motion of Earth relief or clouds and the view of the circumferential horizon of Earth enabled the crew to orient the spacecraft (Molodtsov, 2001)

Mercury flew the first American into space. Like Vostok, it was a single crew spacecraft. There were 06 crewed Mercury flights (Mercury 3, 4, 6-9) from 1961-63. Like Vostok, Mercury stayed in low Earth orbit, till retro-fire and was ballistic through its atmospheric descent (NASA, McDonnell Aircraft Corporation, 1962). Though all flight control tasks were automatic, crew could accomplish three axis attitude control in orbit and during retrofire (Voas, 2008). Mercury had two means of external visual reference: -

- Mercury used a periscope window like Vostok to use Earth as attitude reference. (NASA, McDonnell Aircraft Corporation, 1962).

- Additionally, there was a large observation window like an aircraft windscreen, for external reference. In addition to providing a horizon view, the window permitted celestial navigation using on track stars. Since, the crew preferred the observation window over the periscope, Mercury 9 did away with the periscope (Swenson, Grimwood, & Alexander, 1989). Inability to use the periscope during orbital night and poor illumination due to the system of lenses are mentioned as limitations of the periscope viz a viz the window in (Swenson, Grimwood, & Alexander, 1989) and (Voas, 2008). Periscope and hence bottom view has not featured in any subsequent manned spacecraft of NASA.

There were two manned multi crewed Voskhod flight (1964-65). Voskhod added extravehicular activity to the scope of manned space flight. Manual flight control in Voskhod was the same as Vostok (Tiapchenko, 2003).

Gemini expanded the scope of manned space flight to on-orbit rendezvous/ docking and controlled lift re-entry. There were 10 crewed Gemini flights (1965-66) Gemini was crewed by a complement of two astronauts. Unlike Vostok/ Mercury/ Voskhod where the entire mission could be flown automatically with manual flight control being optional, in Gemini there were several critical flight control tasks that could only be done manually (McDonnell Aircraft Corporation, 1966). These tasks are enumerated below.

- Separation from the launch vehicle and velocity trimming for orbital correction could only be done manually.

- Rendezvous and docking were done manually.

- Achievement and maintenance of de-orbit attitude could only be done manually.

- Several pre-retrofire sequencing commands could be issued only manually. The Human Machine Interface (HMI) provided appropriate situational awareness to the crew to undertake these tasks.

- Manual duplication of automatic retro-fire initiation command.

- Spacecraft attitude during retro-fire could only be controlled manually.

- End of retrofire till descent altitude of 400,000 ft attitude could only be controlled manually using external horizon as reference through observation windows.

There were two observation windows for visual reference (McDonnell Aircraft Corporation, 1966). The windows were not flush on the spacecraft surface but were positioned in a suitably shaped well. It is evident that attitude cues were not by viewing the entire circumference of the Earth (periscope method in Vostok, Mercury, Voskhod) but by using a portion of the visual horizon, like aircraft flying (and Mercury 9).

Apollo expanded the scope of manned space flight lunar orbit and lunar landings. There were 11 manned Apollo missions (1968-72). Only the command module has been discussed here as it was required to perform the de-orbit attitude task that is of our interest. Excluding lunar orbit, landing and takeoff, Apollo's manual flight control tasks were like Gemini. All tasks were automated with the option of manual reversion. Three windows above the control panels provided visual reference for all manual tasks (NASA, 1966).

Soyuz has been in service since 1966. Its present version, Soyuz MS is currently the fastest means to reach ISS. In addition to the manual 3 axis attitude control tasks encountered earlier i.e. to achieve and maintain orbital prograde/ retrograde attitude and achieve and maintain attitude during de-orbit burn. Soyuz also featured manual achievement of electrical power generation attitude. This was done by adjusting the spacecraft's orientation in orbit to point its solar panels towards the sun. A periscope that permitted viewing the entire circumference of the Earth while in orbit was used for visual reference. The periscope display had appropriate markings. A mechanism in the periscope also permitted viewing along the longitudinal axis of the spacecraft for rendezvous and docking, this view was however not used for prograde/ retrograde/ de-orbit attitude tasks. A camera view also supported docking. Solar pointing could be achieved by either view (Soyuz Crew Operations Manual SoyCOM ROP 19, 1999).

Shenzhou the Chinese spacecraft, had its first crewed launch in 2003 and has been operational since then. Shenzhou design is based on the Soyuz. Limited information is available about the spacecraft and its manual flight control. (Yang, 2021) sheds some light on specifications of Shenzhou. Since the spacecraft is based on Soyuz and is being used for transportation to Tianzhou space station, it may be surmised that its manual flight control tasks would be similar to Soyuz. Cockpit images indicate a periscope type display. (Yang, 2021) also mentions that manual docking is done using a docking camera, without any mention of the periscope. The display seemed suitable for providing a view of the Earth's circumference for attitude reference in orbit and to locate the Sun for electrical power generation.

Crew Dragon flew is first crewed mission in 2020 and is currently operational. It is assumed that manual flight control tasks would be typical of manned transportation spacecraft – i.e. on orbit 3 axis attitude control, rendezvous and docking, de-orbit, and re-entry flight path control. Unlike Soyuz, there are no deployable solar panels in the spacecraft obviating the requirement of manual achievement of power generation attitude. Crew Dragon windows don't seem to be located in a manner that provides appropriate visual flying references for de-orbit. Demo 2 mission video indicates an infra-red camera view of a sector of the horizon on the display screen.

Boeing Starliner manned spacecraft under development that has been contracted by NASA. It is yet to fly its first crewed mission. However, uncrewed missions have been flown. Open-source information about manual flight control is scant. Since the spacecraft has been contracted by NASA, it is assumed that its manual flight control tasks would be typical of manned transportation spacecraft. Images available on opensource show a suitably positioned window for visual reference during docking which might also give a 'front' view of the Earth's horizon adequate for Earth referenced orbital attitude tasks.

Orion is a deep space exploration spacecraft under development, that is being developed for NASA. Limited open-source information is available about Orion. However, online images clearly indicate the presence of four horizon viewing windows, two control sticks and two translation controllers.

In summary, the following deductions can be drawn from the literature survey: -

- Manual flight control provision exists in all past, present, and developmental manned spacecraft. On-Orbit flight control includes on-orbit 3 axis orientation/ translation, rendezvous, and docking (where applicable), power generation attitude (where applicable) and de-orbit. Translation controls are included in docking spacecraft.

- Visual reference cues are provided using windows, periscope mechanisms and cameras. Soyuz and Shenzhou prefer to view the entire circumference of the Earth for attitude reference. This reference method lost favour with NASA after Mercury. NASA spacecraft prefer an aircraft windscreen type 'front' view. Front view is required by all docking spacecraft for docking. Soyuz uses a periscope and camera to switch to front view, whereas Shenzhou uses a camera.

Table 2 summarizes on-orbit manual control methods for manned transport spacecraft.

Table-2 List of Viewport Orientations in Different Missions

| S No | Name | External Visual Reference |
|---|---|---|
| 1. | Vostok | Bottom view |
| 2. | Mercury | Bottom + Front view |
| 3. | Voskhod | Bottom view |
| 4. | Gemini | Front view |
| 5. | Soyuz | Front + Bottom view |
| 6. | Apollo | Front view |
| 7. | Space Shuttle | Front view |
| 8. | Crew Dragon | Front view |
| 9. | Shenzhou | Front + Bottom view |
| 10. | CST-100 | Front view |
| 11. | Orion | Front view |
| 12. | Dream Chaser | Front view |

Literature survey reveals front view has obvious advantages for docking spacecraft where the docking port is typically along the front view. However, spacecraft like Soyuz continue to retain bottom view in addition to front view, thereby displaying a preference for bottom view for tasks such as de-orbit that require orienting the spacecraft locally vertical to Earth. Literature survey does not reveal any study that compares these two views for local vertical tasks.

## 3. Proposed Approach

We developed a virtual reality (VR) crew cabin integrated to a three-axis joystick to compare the front and bottom view for a manual de-orbiting task. VR offers an immersive experience of feeling inside a spacecraft. The following sections described the user study in detail.

**Participants:** All countries initiated human spaceflight program by selecting military pilots as crew members while in recent time, NASA and SpaceX opened space travel to civilians. Considering the range of space crews for future missions, six military test pilots and 6 civilians with gaming experience participated in the study. The test pilots had flying experience ranging from 1800 h to 3000 h over a period of 14 to 21 years of military flying.

**Design:** The spacecraft was oriented with an initial configuration of pitch 0º, bank 102º and heading 104º with respect to de-orbit attitude at the start of simulation. In this attitude a portion of Earth was visible in either of the cameras. The pilot was required to achieve de-orbit attitude using the control stick within an accuracy of 1 deg pitch, 1 deg roll and 6 deg yaw. Attitude information (current body rates/ pitch ladder/ heading scale/ bank/ virtual horizon/ ADI) was not displayed to the crew – since this made external view irrelevant. Pilot flying the simulator could manipulate orientation of the spacecraft by demanding rotation about body axes using a control stick.

**Procedure:** Pilots practiced on the simulator till they were comfortable and confident to achieve the desired attitude accuracy with the control stick, front view, and bottom view. Thereafter, data was collected in two simulation runs, one with front view and the other with bottom view in quick succession. The sequence of runs (front/ bottom) was randomized. Pilots were given about 30 min of practice using either view for familiarization. For data collection, the pilot was required to do this task once with front view and once with bottom view. Data was collected only if the task was accomplished successfully, else the data was discarded, and the pilot had to repeat the task.

**Material:** To compare different kinds of on-orbit attitude control methods, a Virtual Reality (VR) simulator was developed on Unity gaming software. The simulator environment included a spherical rotating Earth with centrally attracting gravitational force and a 3-D CAD model of a spacecraft with an initial velocity of 7.6 km/s to keep in at an orbit of 400 km above Earth. The VR model included a bottom view camera with 145 X 145 deg FOV bore-sighted towards the center of Earth in de-orbit attitude and a front view camera with 70 X 70 deg FOV bore-sighted opposite to the velocity vector in de-orbit attitude. Pilot flying the simulator could be provided with either view on the cockpit display.  Pilots had to wear a restrictive device to simulate spacesuit gloves to make the task more realistic.

We used Thrustmaster 16000T three axis control stick was used. Linear rate demand control law was used, wherein the stick neutral position demanded 0 deg/s rate about body axes, rate demand linearly increased with stick deflection till a maximum of 3 deg/s rate in the respective axis at full deflection. The stick being spring loaded to neutral position resulted in null rate demand at the start of simulation.

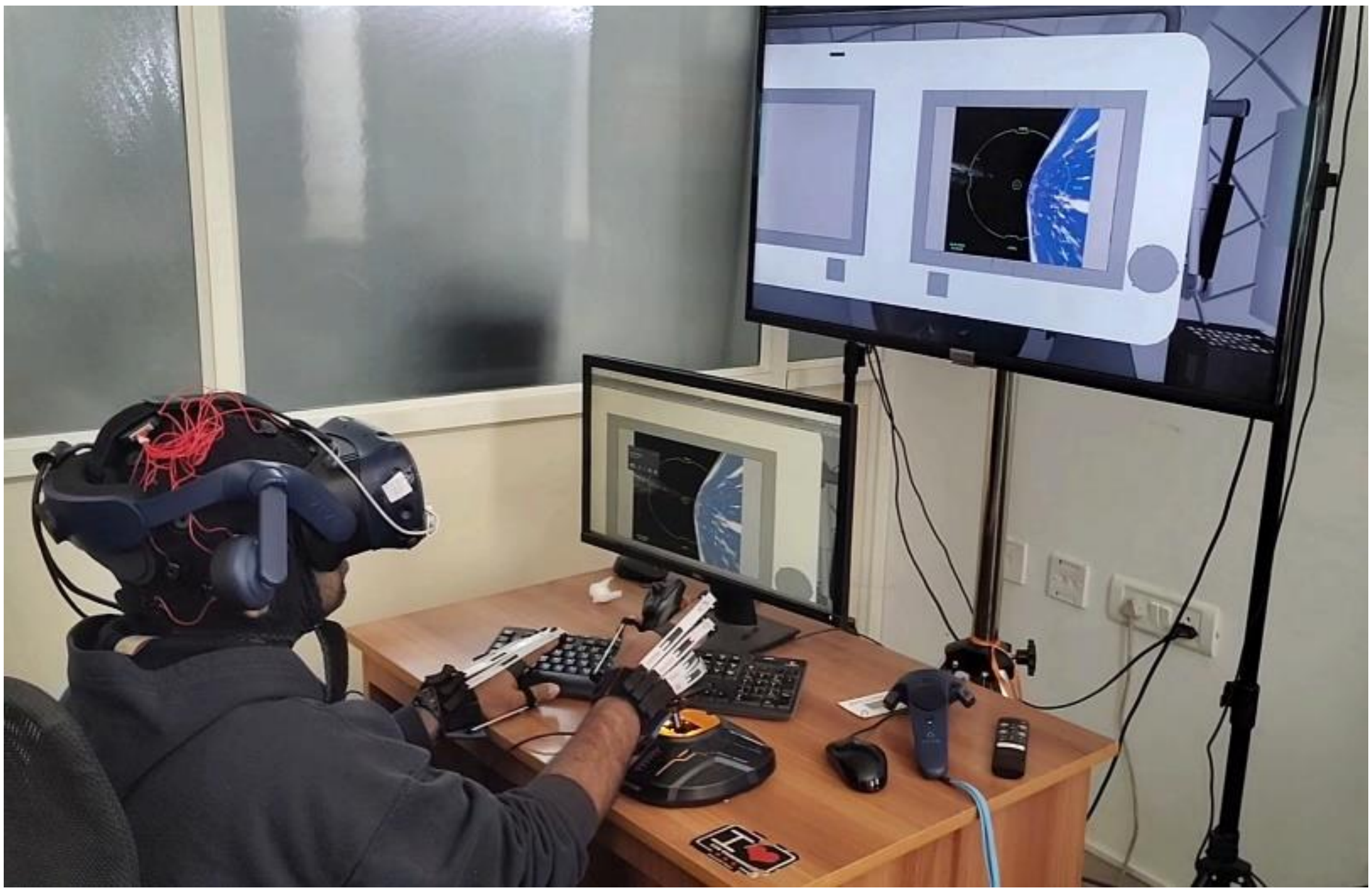

**Figure 1. VR User Study Set Up**

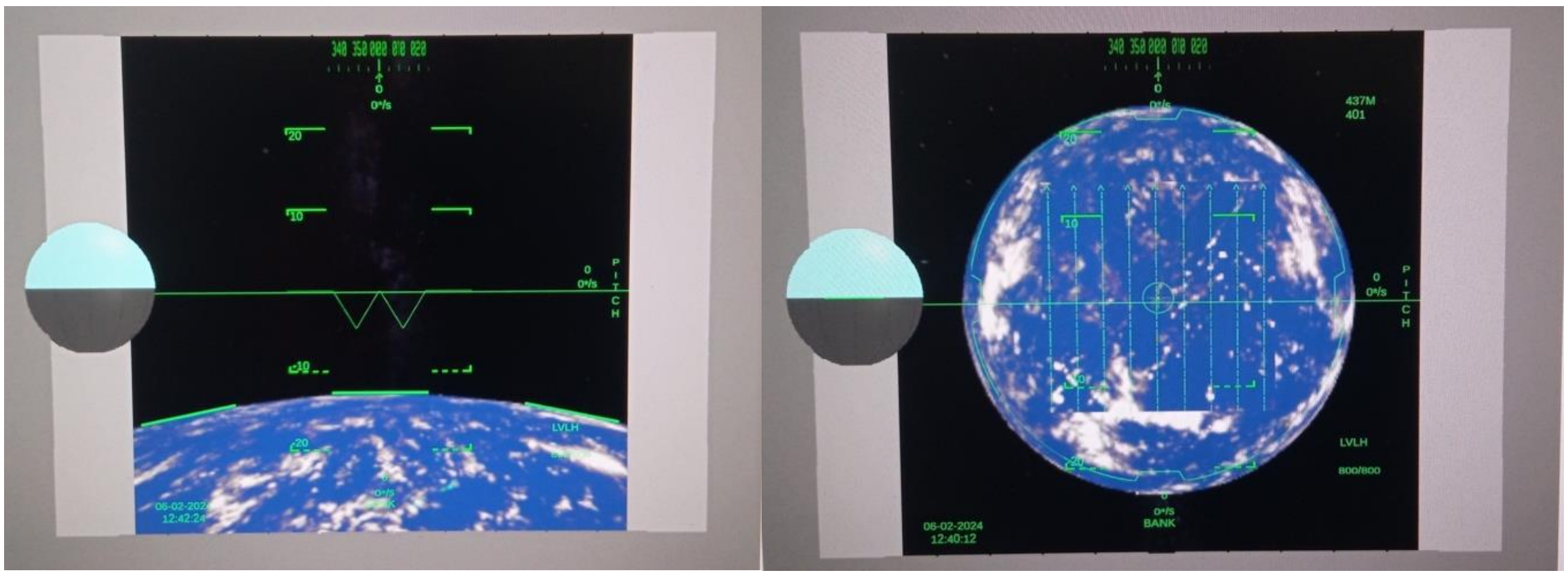

Front View (De-orbit attitude) Bottom view (De-orbit attitude)

**Figure 2. Views Used in the Study**

We used specialized gloves restricting the range of motion of fingers and wrist developed by Cambridge University Engineering Design Centre for motor impairment simulation [Waller 2020].

**Human Factor Analysis:** We used HTC Vive Pro Eye HMD for rendering cockpit view and outside window scenery to the participants. The HMD gives a diagonal FOV of 110° and a resolution of 1440 X 1600 pixels. HTC Vive Pro Eye has an inbuilt eye tracker which is used to record ocular parameters such as x/y gaze direction vectors, left/right pupil positions and pupil size. All required data is synchronized with the spacecraft simulation data and is acquired at 120Hz. We used HTC Vive's SRanipal SDK (VIVE Developers, 2022 and Liu et al., 2022) along with Tobii's XR SDK version 1.8.0 (Tobii XR Devzone, 2022) for eye tracker data recording. Tobii XR SDK uses Gaze-to-Object Mapping (G2OM) algorithm to determine what the user is looking at. We recorded gaze direction from Tobii XR and pupil diameter from SRanipal SDK.
We detected fixations and saccades from gaze direction data through velocity threshold using fixation identification method (Arjun et al., 2021; Mukhopadhyay et al., 2023). Firstly, we computed angle between consecutive gaze direction vectors. We then calculated angular velocity as change in angle divided by the time increment. We identified fixation based on a velocity threshold of 30 degree/sec. If the angular velocity is lesser than the velocity threshold, it is treated as a fixation. Successive fixations are computed as one fixation. Subsequently, we computed fixation rate as the ratio of total number of fixations to total task duration. Mean fixation duration is computed as the ratio of sum of all fixation durations to total number of fixations.
We used Emotiv EPOC Flex-32 channel saline sensor based wireless EEG headset to measure brain's electrical activity. Raw EEG data for each electrode (in µV) is captured at 128Hz. Substantial amount of signal processing and filtering is carried out within the Flex headset to remove the ambient noise and harmonic frequencies (Williams et al., 2020). Flex headset has in-built data pre-processing algorithms which includes a high-pass filter of 0.2 Hz, a low-pass filter of 45 Hz, a notch filter at 50 and 60 Hz, digitization at 1024 Hz and further filtering using a digital 5th order sinc filter. The data is further down sampled to 128 Hz for transmission.
Sensor data is processed into four frequency bands: theta (4-8Hz), alpha (8-12 Hz), beta (16-25Hz) and gamma (25-45 Hz). Emotiv also provides average band power (in µV2/Hz) for each frequency band computed using fast fourier transform (FFT). Before applying FFT, the data is processed through a hanning window of size 256 samples that is slid by 16 samples in each iteration to create a new window (Emotiv, 2023;). Both the raw sensor data and the average power per frequency band for each sensor is stored for each simulation. We used Algorithm 1 to calculate Task Load Index (TLI) and Engagement Index (EI) [Hebbar 2023; Rao 2023].

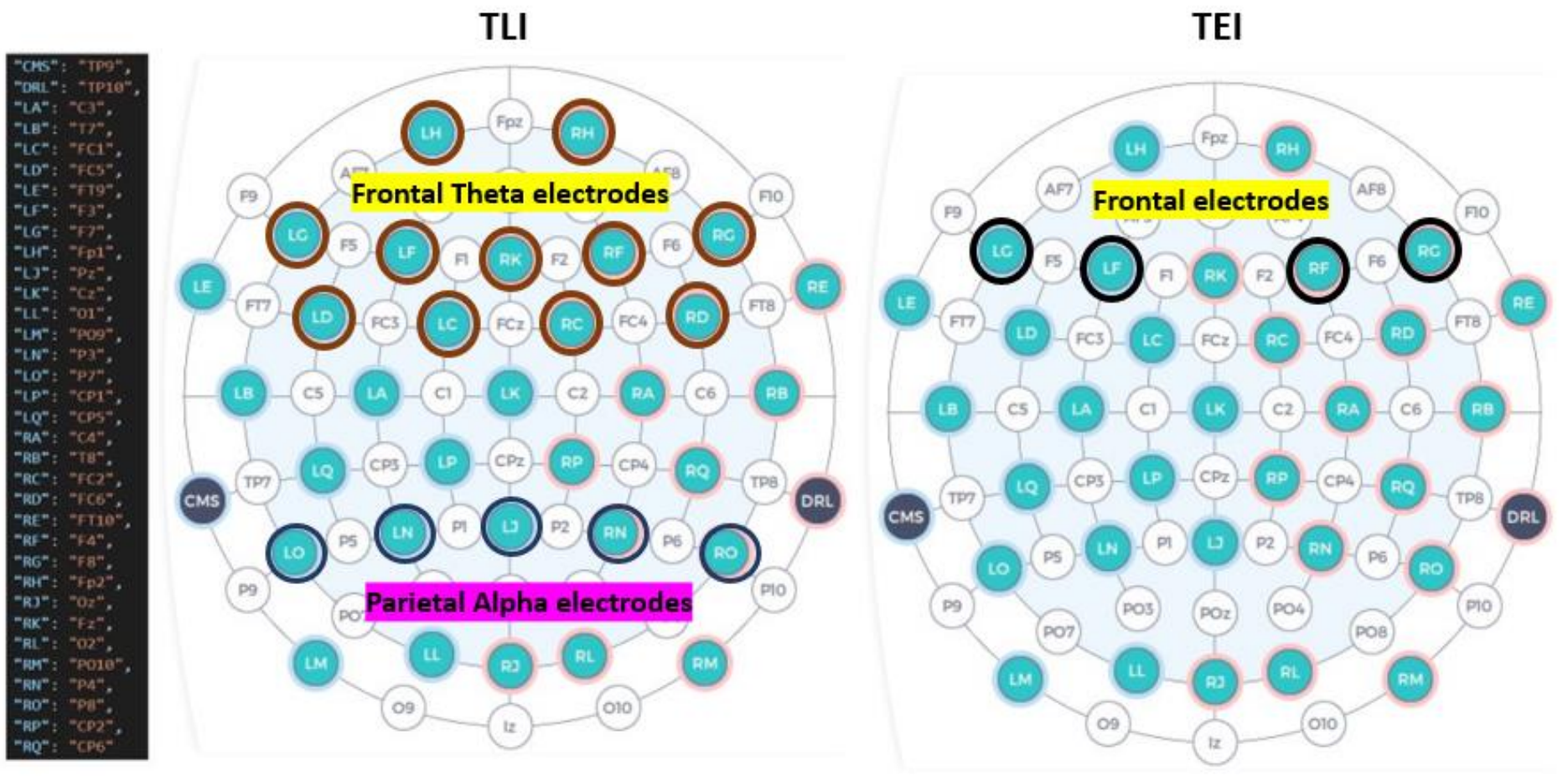

**Figure 3. EEG electrodes considered for TLI and TEI calculation**

---

ALGORITHM 1: Algorithm for Calculating TLI and TEI from EEG Data

---

Input: Raw EEG values for 32 channels

Output: TLI

For each electrode

    Sample raw EEG values through Hanning Window and Fast Fourier Transform

      Band power averaged over 16 samples

For each sample of band power

    Calculate intermediate TLI as mean (theta - Fz, Fp1, Fp2, F3, F4, F7, F8, FC1, FC2, FC5, FC6) / mean (alpha - Pz, P3, P4, P7, P8)

    Calculate intermediate TEI as mean (beta - F3, F4, F7, F8) / {mean (alpha - F3, F4, F7, F8) + mean (theta – F3, F4, F7, F8)}

Compute median of set of intermediate TLI and TEI

Return TLI and TEI corresponding to respective participant and experiment mode

End

---

**Results:** Data was collected on task completion time, fuel consumption and attitude error at 60 Hz. Ocular and EEG data was recorded during the study and analyzed as discussed in the section above. Pilots also filled NASA TLX and IBM SUS forms. After completion of the task, each participant expressed their free opinions about two different viewing strategies.

Task completion time was less in bottom view compared to front view for both groups.
Fuel Consumption had different trends for civilians and pilots. For the front view, it was higher for students than pilots while for bottom view, it was nearly same for both groups. We did not find the difference in task completion times and fuel consumption significant at $p<0.05$.

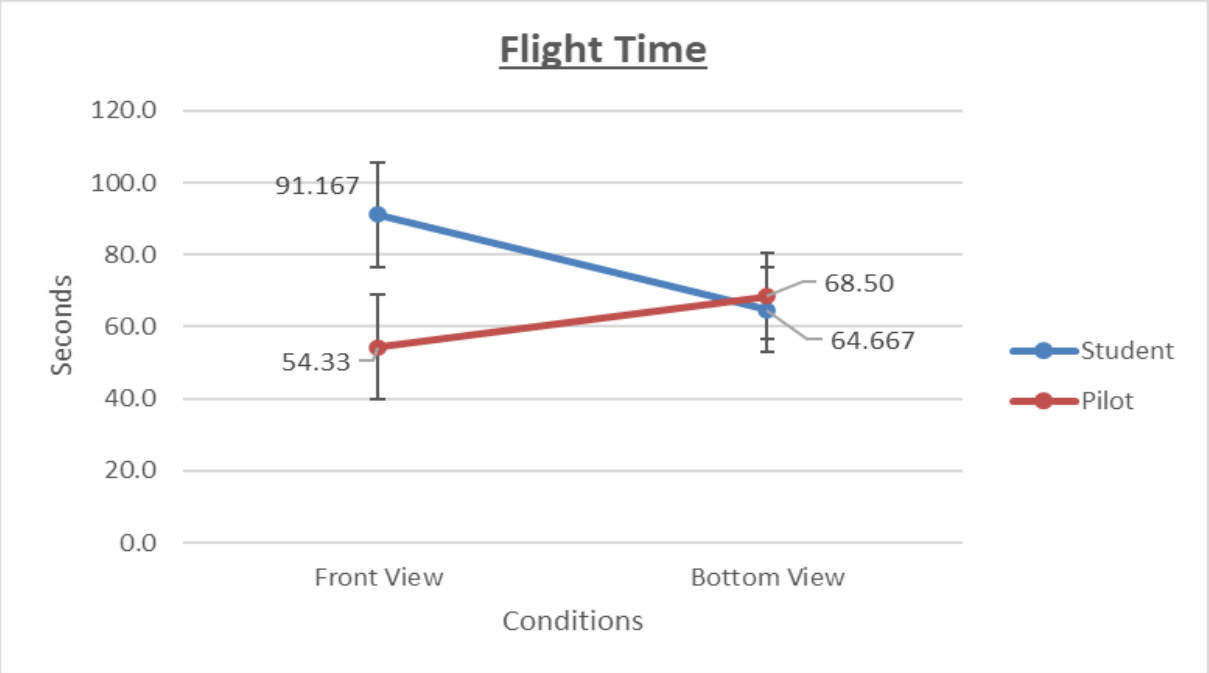

**Figure 4. Comparing the De-orbiting session of flight duration**

## Human Factor Analysis

Eye gaze fixation rate did not show any significant difference with change in external view, however, there was a significant effect of type of external view on saccade velocity ($F(1,10) = 7.508$, $p<0.05$, $\eta^2 = 0.429$), which was lower in bottom view than front view for both group of participants. We found significant interaction effect of view conditions and participants for engagement index $F(1,10) = 3.650$, $p<0.05$, $\eta^2 = 0.267$ while we did not find any significant difference for TLI.

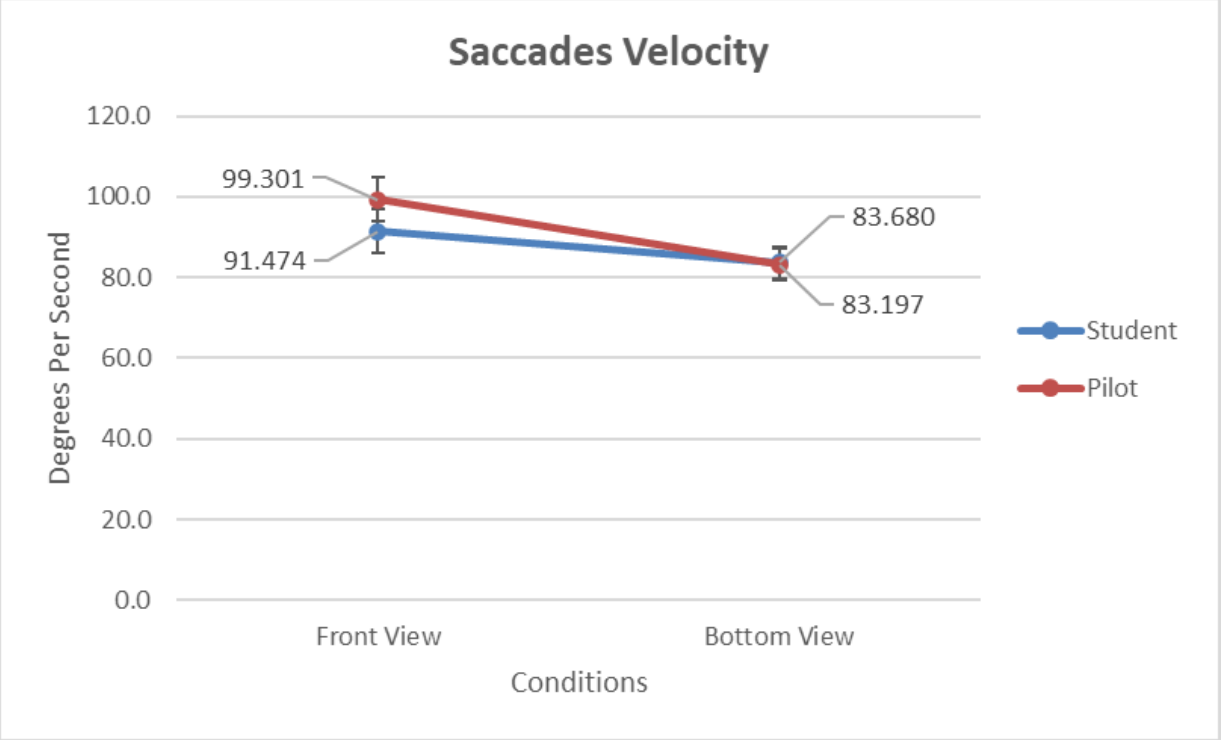

**Figure 5. Comparing the saccades velocity of participants**

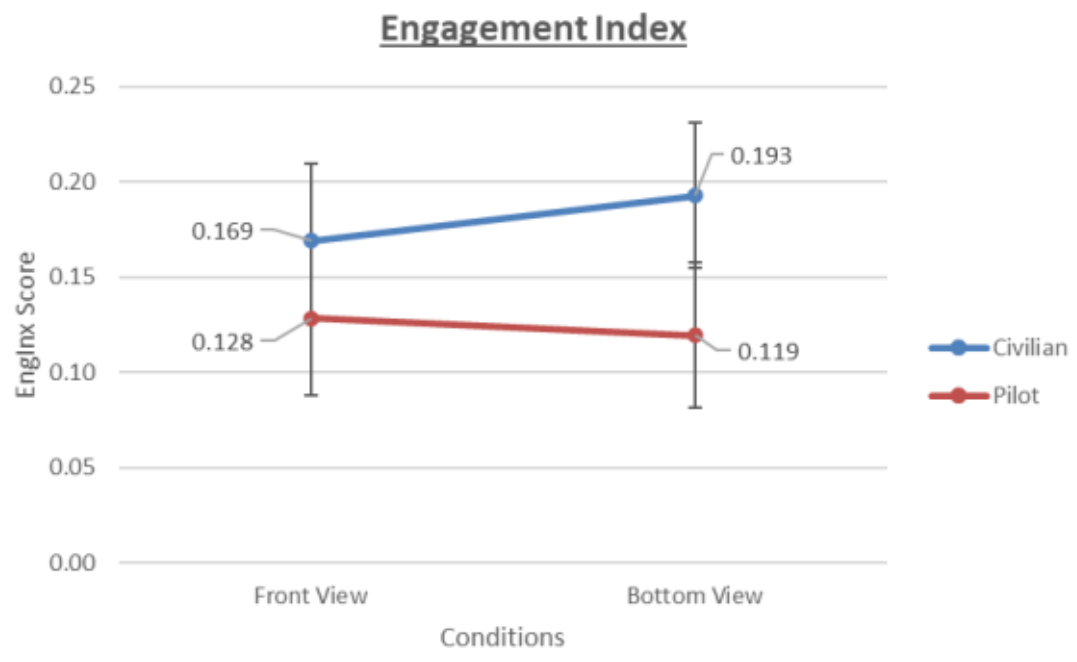

**Figure 6. Comparing the task engagement index of participant**

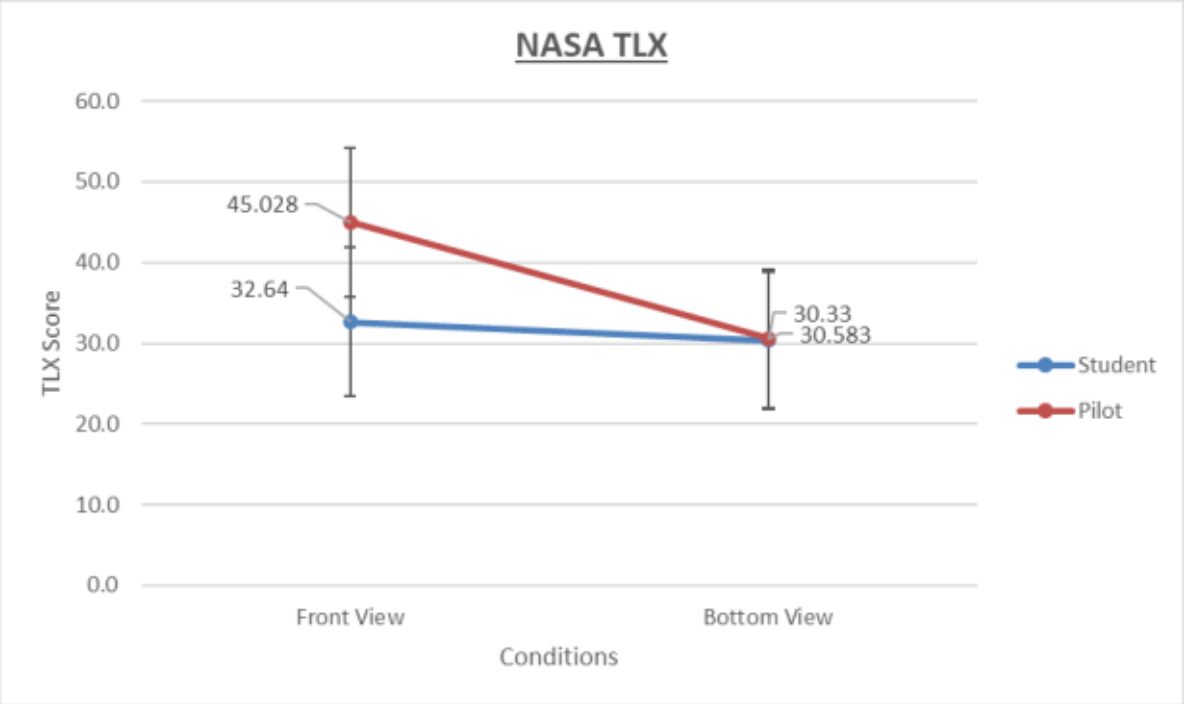

**Figure 7. Comparing the task load index of participants**

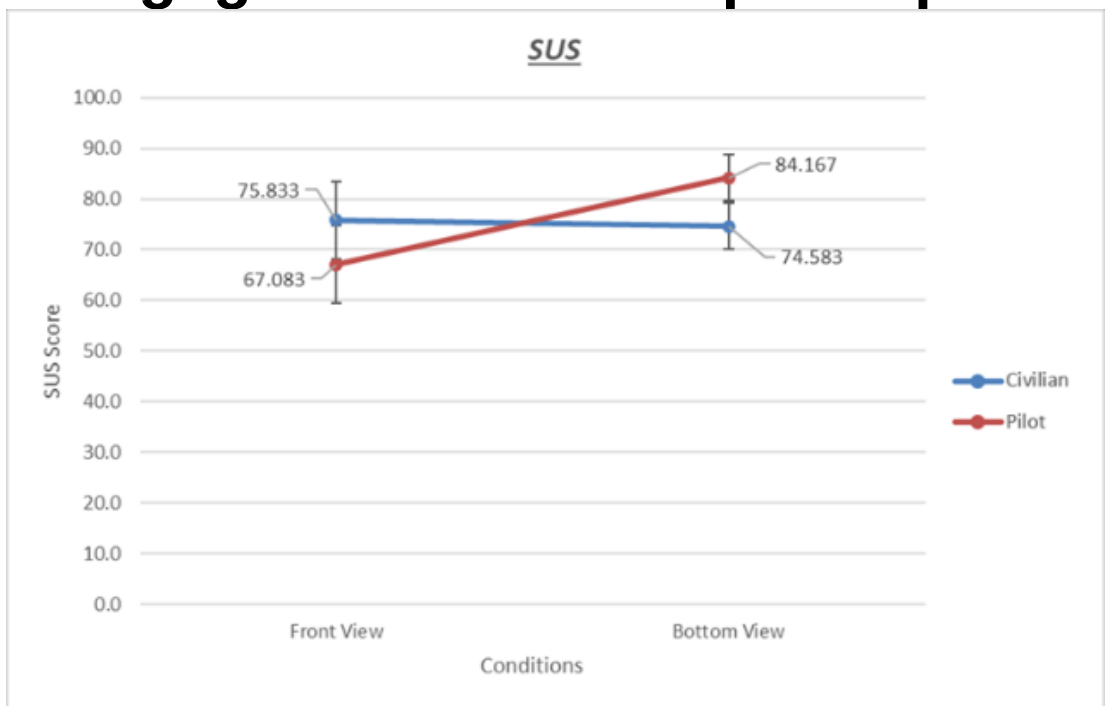

We found significant effect of view conditions for NASA TLX $F(1,10) = 6.466$, $p<0.05$, $\eta^2 = 0.393$, it was lower for bottom view compared to front view. We also found a significant interaction effect of IBM SUS score with pilots preferring the bottom view $F(1,10) = 2.696$, $p<0.05$, $\eta^2 = 0.212$.

**Figure 8. Comparing the IBM SUS of participants**

**Discussion:** Our study demonstrated that a VR based system can work as an excellent training simulator where both test pilots and civilians could learn de-orbiting maneuver and undertake the same in less than 5 minutes with realistic control laws. Either view could be used to de-orbit effectively. Heading correction in front view required holding the inertial attitude steady and observing roll error build up. No such hold and observe was required for bottom view. This explains why the front view resulted in longer task achievement time than bottom view. In terms of fuel consumption, front view resulted in lesser fuel consumption for test pilots than bottom view. In near local vertical attitude, small errors in pitch/ bank showed up more prominently in bottom view than front view. Pilots tended to correct these errors while fine tuning heading resulting in higher fuel consumption in bottom view. Bottom view showed lower cognitive load and higher user preference. This may be explained by the additional effort required for close observation required to determine heading error in front view. indicative of lower cognitive workload with bottom view.
Pilots with prior spacecraft training reported that the task and simulation were realistic. One pilot felt that the display flicker with VR headset caused eyestrain which was not representative of actual spaceflight. However, this flicker was common to both views and hence did not affect this study. It was realized that front view required accurate achievement of near zero bank with respect to local vertical to correctly appreciate heading error. In case of any bank while correcting heading (inadvertent pilot input, control cross-coupling, perturbing torques and so on) it would be very difficult to proceed with further heading correction in front view for de-orbit task. While this was a drawback, front view also provided a design advantage wherein there was no requirement of a zoom camera for heading correction. Also, the front view had obvious advantage for docking as docking ports tend to be front aligned.

## 4. Acknowledgements

The authors thank ISRO scientists Vishal Shukla and Jyoti for assistance in designing the simulator.

## 5. Conclusion

This paper presents a comparison of viewport orientations for de-orbiting maneuver of a spacecraft using a bespoke virtual reality simulator. Based on data from 6 test pilots and an equal number of civilians, the bottom viewport is found to be easier to operate than front viewport for the de-orbiting maneuver. Besides the immediate result, the study also demonstrates use of VR simulator and a set of human factor analysis tools for investigation of man-machine interface of spacecrafts.